\documentclass[conference]{IEEEtran}
\IEEEoverridecommandlockouts

\usepackage{cite}
\usepackage{amsmath,amssymb,amsfonts}
\usepackage{graphicx}
\usepackage{textcomp}
\usepackage{xcolor}
\usepackage{multirow}
\usepackage{threeparttable}
\usepackage{array}
\usepackage{tabularx}
\usepackage{makecell}
\usepackage{booktabs}
\usepackage{bbding}
\usepackage{float}
\usepackage{listings}
\usepackage{amsmath}
\usepackage{xcolor}
\usepackage{enumerate}
\usepackage{makecell}
\usepackage{placeins}
\usepackage{algorithm}
\usepackage[noend]{algpseudocode}
\usepackage{enumerate}
\usepackage{stfloats}
\newtheorem{theorem}{Theorem}

\def\BibTeX{{\rm B\kern-.05em{\sc i\kern-.025em b}\kern-.08em
    T\kern-.1667em\lower.7ex\hbox{E}\kern-.125emX}}
\begin{document}
\title{UREM: A High-performance Unified and Resilient Enhancement Method for Multi- and High-Dimensional Indexes
\thanks{ \textsuperscript{*} Corresponding author.}}

\author{
\IEEEauthorblockN{Ming Sheng\IEEEauthorrefmark{2}, Shuliang Wang\IEEEauthorrefmark{2}\IEEEauthorrefmark{1}, Yong Zhang\IEEEauthorrefmark{3}\IEEEauthorrefmark{1}, Yi Luo\IEEEauthorrefmark{2}\IEEEauthorrefmark{1}, Xianbo Liu\IEEEauthorrefmark{2}, Zeming Li\IEEEauthorrefmark{2}}
\IEEEauthorblockA{\IEEEauthorrefmark{2} Beijing Institute of Technology, Beijing, China \\
\IEEEauthorrefmark{3} Tsinghua University, Beijing, China \\
\IEEEauthorrefmark{2} \{shengming, slwang2011, luoyi, liuxianbo2024, lizeming\}@bit.edu.cn, \IEEEauthorrefmark{3} zhangyong05@tsinghua.edu.cn}
}

\setlength{\abovecaptionskip}{0pt}
\setlength{\belowcaptionskip}{0pt}
\setlength{\parskip}{0pt}

\abovedisplayshortskip=0pt
\belowdisplayshortskip=0pt
\abovedisplayskip=0pt
\belowdisplayskip=0pt

\maketitle

\begin{abstract}
Numerous multi- or high-dimensional indexes with distinct advantages have been proposed on various platforms to meet application requirements.
To achieve higher-performance queries, most indexes employ enhancement methods, including structure-oriented and layout-oriented enhancement methods. 
Existing structure-oriented methods tailored to specific indexes work well under static workloads but lack generality and degrade under dynamic workloads. 
The layout-oriented methods exhibit good generality and perform well under dynamic workloads, but exhibit suboptimal performance under static workloads. 
Therefore, it is an open challenge to develop a unified and resilient enhancement method that can improve query performance for different indexes adaptively under different scenarios.

In this paper, we propose UREM, which is the first high-performance \underline{U}nified and \underline{R}esilient \underline{E}nhancement \underline{M}ethod designed for both multi- and high-dimensional indexes, capable of adapting to different scenarios. 
Specifically, UREM 
(1) can be uniformly applied with different indexes on various platforms; 
(2) enhances the query performance of indexes by layout optimization under static workloads;
(3) enables indexes to stabilize performance when queries shift through partial layout reorganization. 
We evaluate UREM on 20 widely used indexes. 
Experimental results demonstrate that UREM improves the query performance of multi- and high-dimensional indexes by up to 5.73$\times$ and 9.18$\times$ under static workloads, and by an average of 5.72$\times$ and 9.47$\times$ under dynamic workloads. Moreover, some traditional indexes enhanced by UREM even achieve performance comparable to or even surpassing that of recent advanced indexes.
\end{abstract}


\section{Introduction}
Multi- or high-dimensional queries play a vital role in various tasks such as multimodal data exploration~\cite{MMDexploration}, data mining~\cite{MMDmining}, and multimodal knowledge graphs~\cite{MMGraph}. 
Moreover, these queries are also essential for ensuring the performance of Large Language Models (LLM)-based applications, including semantic integration~\cite{SemanticI}, Retrieval-Augmented Generation (RAG) ~\cite {Chameleon, InteractiveRAG}, and personalized Q\&A~\cite{PersonQ&A}. Among these queries, multi-dimensional queries mainly include Predicate-Based Filtering (PBF), which filters data according to query predicate ranges, and Approximate Nearest Neighbor Search (ANNS), which identifies data similar to a given query ~\cite{learnedSUR}. In contrast, high-dimensional queries are primarily focused on ANNS in high-dimensional spaces~\cite{KNNevaluating}.
\begin{table*}[htbp]
  \caption{Indexes and their corresponding enhancement methods.}
  \label{Trelatedwork}
  \begin{threeparttable} 
  \begin{tabularx}{\textwidth}{m{6.2cm}<{\centering}|m{1.2cm}<{\centering}|m{0.9cm}<{\centering}|m{1.4cm}<{\centering}|m{0.95cm}<{\centering}|m{1cm}<{\centering}|m{1.4cm}<{\centering}|m{1.6cm}<{\centering}}
    \toprule
    \multirow{2}{*}{Index} & \multirow{2}{*}{Index Type}\tnote{a} & \multirow{2}{*}{Platform}\tnote{b} & \multirow{2}{*}{Query Type} & \multicolumn{4}{c}{Enhancement Method}  \\
    \cline{5-8} & & & & Type\tnote{c} & Workload & Data-driven & Query-driven  \\
    \midrule
    Z-order~\cite{zorder}, ZM~\cite{ZM}  &  \multirow{5}{*}{Multi-} & \multirow{6}{*}{RDB} & PBF &\multirow{3}{*}{structure-} & \multirow{4}{*}{Static}& \Checkmark & \XSolidBrush \\
    \cline{1-1} \cline{4-4} \cline{7-8}
    Flood~\cite{Flood}, Tsunami~\cite{Tsunami}, BLAEQ~\cite{BLAEQ}, Qd-tree~\cite{Qd-tree} & & & PBF &  & &\Checkmark & \Checkmark  \\ \cline{1-1}\cline{4-4} \cline{7-8}
    R*-tree~\cite{r_tree2}, ML~\cite{Ml-index}, LISA~\cite{LISA}, LIMS~\cite{LIMS} & & & PBF, ANNS  &  & & \Checkmark & \XSolidBrush  \\ \cline{1-1}\cline{4-5} \cline{7-8}
    Fineblock~\cite{fine-grained}, Pando~\cite{pando}& & & PBF  & \multirow{3}{*}{layout-} &  & \Checkmark & \Checkmark  \\ \cline{1-1}\cline{4-4} \cline{6-8}
    OREO~\cite{OREO}, SAT~\cite{SAT}, MTO~\cite{MTO}  &  &  & PBF &  & \multirow{3}{*}{Dynamic} & \Checkmark & \Checkmark \\ \cline{1-2} \cline{4-4} \cline{7-8}
    Cracking~\cite{cracking}, DEPA~\cite{depa}, MetaAdaptive~\cite{adaptiveindex}  & One- &  & PBF &  & & \Checkmark & \Checkmark \\
    \hline
    E$^{2}$LSH~\cite{E2LSH}, IVFADC~\cite{IVFADC}, HNSW~\cite{HNSW}, HAMG~\cite{hamg}, Finger~\cite{Finger} & \multirow{3}{*}{High-} & \multirow{2}{*}{VDB} & ANNS & \multirow{3}{*}{structure-} & \multirow{3}{*}{Static} & \Checkmark & \XSolidBrush  \\  \cline{1-1}\cline{4-4} \cline{7-8}
    FLEX~\cite{FLEX}, DB-LSH~\cite{DB_LSH} & & & ANNS & & & \Checkmark & \Checkmark\\ \cline{1-1} \cline{3-4} \cline{7-8}
    MQRLD~\cite{mqrld} & & DL & PBF, ANNS & & & \Checkmark & \Checkmark\\ \midrule 
    Index + UREM (ours)  & Unified & Unified &  Unified & layout- & Adaptive & \Checkmark & \Checkmark \\
  \bottomrule
\end{tabularx}
\begin{tablenotes}
\footnotesize
\item[a] "Multi-" refers to Multi-dimensional data, "One-" refers to One-dimensional data, and "High-" refers to High-dimensional data.
\item[b] "RDB" refers to Relational Database, "VDB" refers to Vector Database, and "DL" refers to Data Lake.
\item[c] "structure-" refers to structure-oriented enhancement method, and "layout-" refers to layout-oriented enhancement method.
\end{tablenotes}
\end{threeparttable}
\end{table*}

To support multi- or high-dimensional queries, constructing efficient and effective indexes is a central research task, especially under diverse application scenarios. 
Existing studies typically employ two types of indexes, which are multi-dimensional indexes and high-dimensional indexes. 
These indexes are designed to provide high query efficiency (and accuracy) under static workloads, or ensure stable performance under dynamic workloads.

To achieve these objectives, existing indexes often employ enhancement methods, as summarized in Table~\ref{Trelatedwork}, which include \textbf{structure-oriented enhancement methods} and \textbf{layout-oriented enhancement methods}. Most existing indexes rely on structure-oriented enhancement methods that construct reliable customized index structures by leveraging data distribution (data-driven)~\cite{r_tree2, ZM, Ml-index, zorder, LISA} or historical queries (query-driven)~\cite{Flood, LIMS, Tsunami}. 
Due to the high customization, these methods often demonstrate superior query efficiency (and accuracy) under static workloads. However, they lack generality, and their performance often becomes unstable when faced with dynamic queries.

Other indexes improve query performance through layout-oriented enhancement methods, which optimize data layouts by capturing data characteristics or query information~\cite{OREO, depa, SAT, MTO}. These methods exhibit good generality and demonstrate significant advantages under dynamic workloads, since most of them are independent of the upper-level index structures and can dynamically reorganize data layouts to reduce index reconstruction overhead.
However, some of these methods lack customization and often yield suboptimal performance under static workloads. Moreover, these methods are primarily designed for PBF queries on low-dimensional data, which limits their applicability to ANNS queries over multi- or high-dimensional data.

In this paper, we propose UREM,  a \textbf{U}nified and \textbf{R}esilient \textbf{E}nhancement \textbf{M}ethod that firstly improves the query performance of multi- and high-dimensional indexes adaptively in different scenarios.
First, various scenarios demand that an enhancement method should be seamlessly applicable to different indexes across multiple platforms (e.g., relational and vector databases), demonstrating generality. However, structure-oriented enhancement methods that rely on customized index structures inherently lack such generality. Therefore, we design UREM as a layout-oriented enhancement method by decoupling it from the upper-layer index. 
Second, it is essential for indexes to achieve superior query performance under static workloads due to the accumulation of large volumes of data and historical queries. However, existing layout-oriented enhancement methods often suffer from limited customization and consequently yield suboptimal performance under static workloads. By conducting a deep offline analysis of data characteristics and query patterns, UREM enhances the query efficiency (and accuracy) of existing indexes under static workloads.
Third, in many application scenarios, queries are continuously shifting, making it crucial to stabilize query performance under dynamic workloads. UREM adaptively reorganizes and switches data layouts in response to workload variations, thereby improving the stability of existing indexes under dynamic workloads. Our main contributions are summarized as follows: 

\begin{itemize}
\item We are the first to present a unified and resilient enhancement method, UREM, which can be seamlessly applied to various indexes across different platforms. It is a layout-oriented enhancement method that decouples from the upper-layer index.
\item A layout optimization strategy under static workloads is proposed to improve query efficiency (and accuracy). It clusters similar data and data frequently queried together through a data- and query-driven adjustment strategy.
\item A partial layout reorganization mechanism under dynamic workloads is proposed to reduce performance fluctuations of existing indexes.
It dynamically generates and selects the optimal partial layout to stabilize query performance.
\item We evaluate UREM on 20 widely used indexes. Results demonstrate that indexes enhanced by UREM achieve orders-of-magnitude improvements in query performance under both static and dynamic workloads. Moreover, some traditional indexes enhanced by UREM even achieve performance comparable to or even surpassing that of recent advanced indexes.

\end{itemize}

The paper is constructed as follows: Section~\ref{back&moti} presents the background and motivation. Section~\ref{overview} gives an overview of UREM, followed by layout optimization under static workloads in Section~\ref{DeeplyDataOrganization} and partial layout reorganization under dynamic workloads in Section~\ref{PartialOrganization}. Section~\ref{experiments} reports experimental results. Section~\ref{discussion} provides discussion. Section~\ref{relatedwork} reviews related work, and Section~\ref{conclusion} concludes the paper.

\section{Background and Motivation}\label{back&moti}
\subsection{Scenario Characteristics and Motivation}

Different query scenarios, including diverse platforms, query types, and workloads, exhibit distinct query requirements for various indexes. Consequently, the design of an enhancement method is motivated by the following characteristics:

1) \textit{Diverse Indexes and Platforms.} As shown in Table~\ref{Trelatedwork}, existing indexes can be categorized into two types.
The first is multi-dimensional indexes, suited for data with relatively low dimensionality, such as tabular and spatial data, with 2 to 10 dimensions~\cite {Multisurvey}. 
These indexes, mostly used in relational databases, perform well for PBF~\cite{ZM, Ml-index, LISA, RSMI, Flood, Tsunami} or ANNS~\cite{Qd-tree, r_tree2}. 
The second is high-dimensional indexes, which are applied to data such as images and audio, typically represented by high-dimensional feature vectors. 
These indexes are mostly used in vector databases and designed to support ANNS~\cite{HNSW, IVFADC, E2LSH, DB_LSH}. 

2) \textit{Complex Static Workloads.}
In real-world applications, substantial volumes of data and historical queries have been accumulated, which often exhibit complex distributions~\cite{efficient, glove}. For example, logs from university course management systems reveal that different courses are offered each semester, while students from different majors demonstrate distinct query behaviors and preferences~\cite{skewdata}. Therefore, enabling indexes to be data- and query-driven is essential for achieving efficient (and accurate) queries. 

3) \textit{Various Dynamic Workloads.} Recently, several studies have focused on dynamic workloads, as queries in real-world scenarios are constantly shifting. For example, on social media, trending topics change frequently and often lead to sudden surges in user traffic~\cite{weibo}. This necessitates rapid and resilient adaptation of data layouts or indexes to ensure service performance~\cite{A-Tune-Online}. 

Therefore, it is beneficial for improving the performance of downstream tasks by developing a method that can uniformly and resiliently enhance query performance across different scenarios.

\subsection{Existing Enhancement Methods}
Existing enhancement methods can be categorized into structure-oriented and layout-oriented enhancement methods. Most prior studies have focused on structure-oriented enhancement methods. However, these methods are often highly customized and thus difficult to generalize across diverse scenarios. In contrast, layout-oriented enhancement methods can be designed to decouple from both upper-level indexes and lower-level platforms, endowing them with strong potential for unified enhancement.

For static workloads, layout-oriented enhancement methods leverage workload information (e.g., query predicates) and data distribution to design layouts tailored to specific datasets, thereby achieving strong data skipping performance~\cite{MTO, fine-grained, pando}. However, they are difficult to extend discussions on sustaining performance under dynamic queries. 

Subsequent studies have proposed several strategies used in dynamic workloads. Some studies propose repartitioning and merging as queries increase until the layout converges~\cite{depa, cracking, adaptiveindex}. However, these methods overlook handling sudden skewed queries after convergence. Some other studies leverage deep learning models to determine optimal layout~\cite{MTO}, but they result in high reorganization overhead. Furthermore, other works periodically reorganize layouts based on predefined thresholds, ensuring that the current layout remains optimal~\cite{OREO, SAT}. Comparing among these works, one work~\cite{OREO} demonstrates stronger generalization, but it only supports PBF on multi-dimensional data and lacks discussion on static workloads. 

Overall, although existing layout-oriented enhancement methods can be applied to static or dynamic workloads, they lack the flexibility to be effectively adapted to both. Furthermore, most of these methods are tailored for multi-dimensional PBF queries, which makes them difficult to be extended to ANNS on multi- or high-dimensional data.

\subsection{Challenges of Existing Enhancement Methods}
Overall, our study highlights three major challenges:

\textbf{1) \textit{How to enhance the query performance of existing indexes with a unified method?}} Most enhancement methods are limited to specific scenarios. For example, the majority of structure-oriented enhancement methods are designed for specific indexes, while most layout-oriented enhancement methods support only PBF queries on multi-dimensional data, making cross-index integration difficult. Furthermore, indexes like R*-tree~\cite{r_tree2} and HNSW~\cite{HNSW} do not simultaneously utilize query patterns and data characteristics to improve query performance, even though they still have the potential to benefit from other enhancement methods.

\textbf{2) \textit{How to improve the query efficiency (and accuracy) of existing indexes under static workloads?}}
Under static workloads, the primary goal of existing multi-dimensional indexes is to achieve more efficient queries. Moreover, for high-dimensional indexes, achieving more efficient and accurate queries remains a challenge. Therefore, it is a key objective to improve the query efficiency (and accuracy) of existing multi- and high-dimensional indexes under static workloads.

\textbf{3) \textit{How to stabilize query performance of existing indexes under dynamic workloads?}}
Most existing indexes are designed based on fixed data distributions and historical queries. When applied to dynamic queries, their performance often fluctuates. Therefore, it is still a significant challenge to stabilize the query performance of multi- and high-dimensional indexes under dynamic workloads. 

\begin{table}[t!]
\scriptsize
  \caption{Notations.}
  \label{Notations}
  \begin{tabular}{ll}
    \toprule
    Notation & Description  \\ \midrule 
    $D$ & A set of data points with cardinality $N_d$ and dimensionality $d$\\
    $Q$ & A set of query points with cardinality $N_q$ and dimensionality $d$\\
    $T$ & The vertical concatenation matrix of $D$ and $Q$\\
    $T_{\text{trans}},D_{\text{trans}}$ & $T$ and $D$ after feature-based transformation\\
    $T_{\text{adj}},D_{\text{adj}}$ & $T$ and $D$ after data- and query-driven data adjustments\\
    $D_{\text{opt}}$ &The optimized data layout after partitioning process\\
    $||a,b||$ & The Euclidean distance between $a$ and $b$\\
    $\text{KEY}(a)$ & The sorting value of point $a$ under static workloads\\
    $\text{DIS}(a)$ & The sorting value of point $a$ under dynamic workloads\\
    $W_i$ & The $i$-th sliding windows\\
    $Q_{W_i}$ & The queries within the $i$-th sliding windows\\
    $L$ & The candidate layout pool \\
    $l_i$ & A partial layout with cardinality $N_l$ and dimensionality $d$ \\
    $c(q,l_i)$ & The estimated query cost of query $q$ on the partial layout $l_i$\\
    $C_i$ & The cost vector of $l_i$ within a sliding window\\
    $ctotal_i$ & The cumulative query cost of $l_i$\\
    $\gamma, \beta, \varepsilon, \alpha$ & Parameters \\
  \bottomrule
\end{tabular}
\end{table}

\begin{figure*}[tb]
  \centering \includegraphics[width=0.9\linewidth]{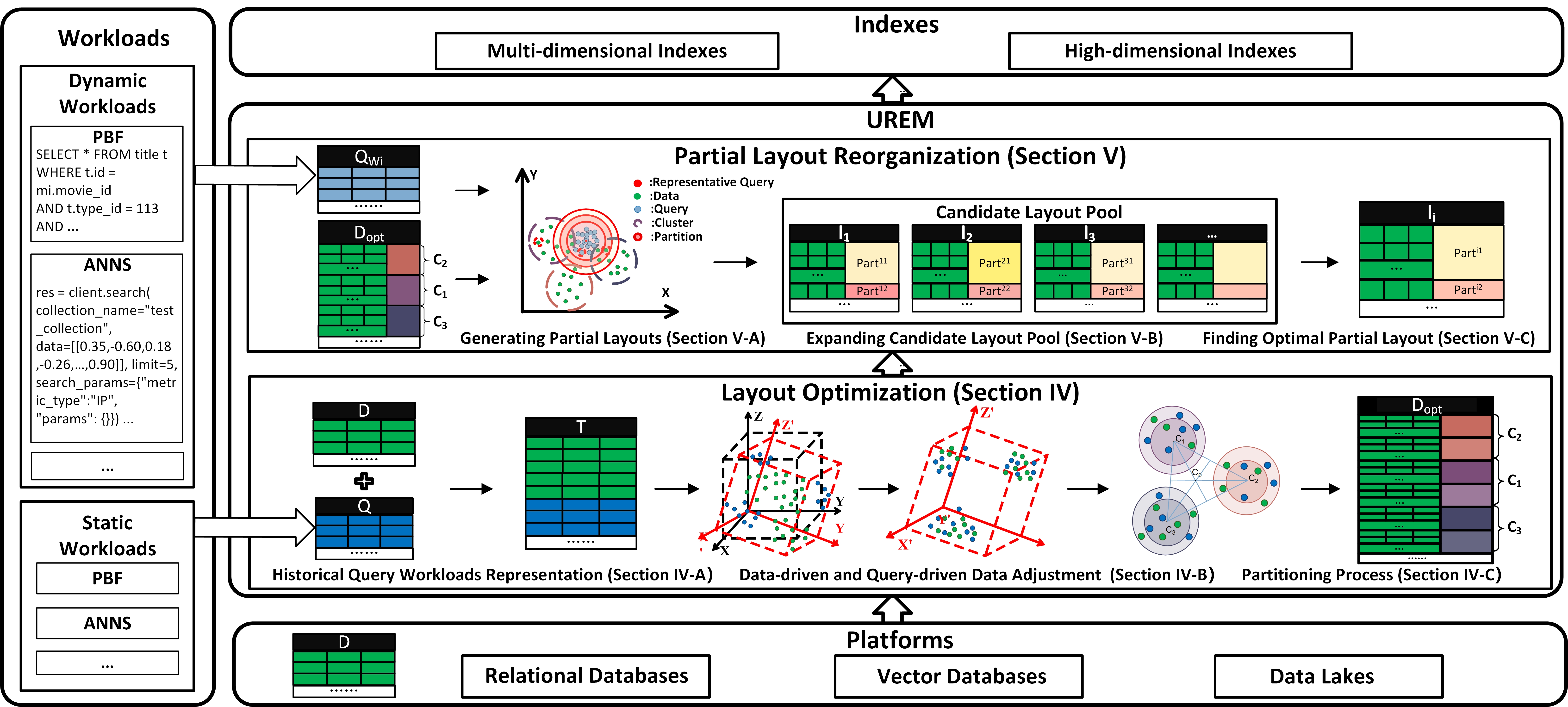}
  \caption{UREM overview.}
  \label{framework}
\end{figure*}

\section{Overview}\label{overview}

UREM is a layout-oriented enhancement method applied before index construction, designed with strong decoupling from underlying platforms, workloads, and index structures, as shown in Figure~\ref{framework}. It serves as a unified method that can be widely used on different platforms such as relational databases, vector databases, and data lakes. Moreover, by leveraging historical or dynamic queries, UREM enables more effective and resilient layout optimization tailored to both static and dynamic workloads. Various types of indexes, including both multi- and high-dimensional indexes, can be built based on UREM to support higher-performance queries. The design of UREM includes the following three features:

\vspace{0.3\baselineskip} 
\noindent \textbf{Unified Integration with Different Indexes on Various platforms.} 
UREM can be integrated with different indexes across various platforms and supports query performance enhancement under both static and dynamic workloads, since it is a layout-oriented enhancement method. Experimental results demonstrate that UREM can be seamlessly combined with 20 commonly used indexes across diverse scenarios (Section~\ref{experiments}).

\vspace{0.3\baselineskip} 
\noindent \textbf{Query Performance Improvement by Layout Optimization under Static Workloads.} 
Under static workloads, UREM offline deeply analyzes data characteristics and query patterns to optimize data layout. It first constructs a unified representation of data and historical queries (Section~\ref{WorkloadsLearning}), then adjusts the data by gathering data that is similar or frequently queried simultaneously, thereby bringing it closer together in hyperspace (Section~\ref{DataAdjustment}). Finally, it partitions the data to produce an optimized layout (Section~\ref{PartitioningProcess}), thereby enhancing query performance.

\vspace{0.3\baselineskip} 
\noindent \textbf{Query Performance Resilience by Partial Layout Reorganization under Dynamic Workloads.} Under dynamic workloads, UREM analyzes query results under continuously shifting queries to enable dynamic layout reorganization and switching. It generates partial layouts to respond to dynamic workloads (Section~\ref{GeneratingPartialReorganization}) and manages a pool of candidate partial layouts (Section~\ref{ExpandingPool}). During query processing, UREM efficiently selects the optimal partial layout from the pool in parallel (Section~\ref{FindingOptimal}) and flexibly executes queries across partial and full layouts (Section~\ref{OnQueryProcess}), thereby ensuring efficient, accurate, and stable query performance.

The following sections present the details of UREM, with commonly used notations summarized in Table~\ref{Notations}.

\section{Layout Optimization}\label{DeeplyDataOrganization}

\begin{figure*}[h]
  \centering
\includegraphics[width=\linewidth]{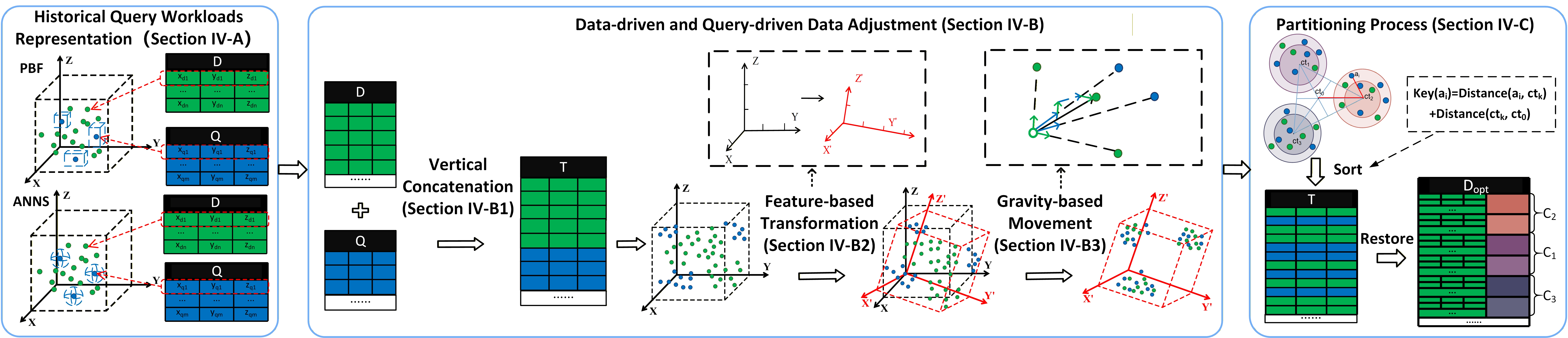}
  \caption{Layout optimization.}
  \label{offlineframework}
\end{figure*}


\subsection{Historical Queries Representation}\label{WorkloadsLearning}

To enhance query performance, UREM adopts a lightweight and effective representation method that enables the consistent representation of all queries and data in a uniform dimensional format.
Existing PBF queries primarily filter records based on one or more range predicates. 
As shown in Figure~\ref{offlineframework}, these queries can be viewed as hypercubes or hyperrectangles within the hyperspace, where the data can be treated as points. 
In this paper, UREM transforms each query into a point by selecting the geometric center of its hypercube or hyperrectangle. 
For predicates with a query range, we choose the median of the range as the representation. 
For predicates without a query range, we use the expected value of all data points satisfying the predicate as its representation. 
ANNS queries are often extended to $k$-ANNS.
Geometrically, these queries can be viewed as a hypersphere. Similarly, UREM abstracts each query by its center point $q$, as illustrated in Figure~\ref{offlineframework}.

\subsection{Data- and Query-driven Data Adjustments}\label{DataAdjustment}
A well-optimized data layout plays a crucial role in minimizing index scans and enhancing query performance. 
In this study, the data undergoes a data- and query-driven adjustments process, which groups data points that are either similar or frequently accessed together, consisting of three main steps: vertical concatenation, feature-based transformation, and gravity-based movement.

\subsubsection{Vertical Concatenation}\label{VerticallyConcatenate}
To enable the unified consideration of both data characteristics and query patterns, UREM vertically concatenates the historical query points $Q$ with the data points $D$ since their representation are in the same dimensionality, constructing a matrix $T$, as illustrated in Figure~\ref{offlineframework}. 

\subsubsection{Feature-based Transformation}\label{ImprovedTransformation}

UREM reshapes $T$ according to the importance and value distribution of each dimension. 
To achieve this, it constructs two matrices, the dimension importance matrix $A_c$ and the data frequency distribution matrix $A_r$. 
In our work, $A_c$ is derived using the hyperspace transformation~\cite{mqrld}, encoding the relative importance of each dimension.
To obtain $A_r$, UREM builds histograms for each dimension in $T$ and derives the matrix through statistical analysis of their value distributions.
The computation proceeds in two steps:

\vspace{0.3\baselineskip}

\noindent \textbf{Step 1:} For $j$-th dimension, we compute the optimal bin width $wid_j$ using the Freedman-Diaconis rule~\cite{rule} to guide the histogram construction for the value distribution:
\begin{equation}
wid_j = 2 \cdot \mathrm{IQR}(x_j) \cdot N^{-1/3}
\end{equation}
where $\mathrm{IQR}(x_j)$ is the interquartile range of values along $j$-th dimension, and $N$ denotes the total number of data points in $T$.

\vspace{0.3\baselineskip} 
\noindent \textbf{Step 2:} Based on the computed bin width, we determine that $b_j$ bins should be constructed for the $j$-th dimension in the histogram:
\begin{equation}
b_j = \left\lceil \frac{\max(x_j) - \min(x_j)}{wid_j} \right\rceil + c_j
\end{equation}
where $\max(x_j)$ and $\min(x_j)$ denote the maximum and minimum values along $j$-th dimension . 
The constant $c_j$ is added to ensure that the total number of bins in each dimension is divisible by the dimensionality $d$, allowing every $b_j/d$ consecutive bins to be merged, thereby forming the data frequency distribution matrix $A_r \in \mathbb{R}^{d \times d}$. 
Therefore, for each $a_{ij} \in A_r$, its value corresponds to the cumulative frequency from the $((i-1) \cdot b_j/d + 1)$-th bin to the $(i \cdot b_j/d)$-th bin in the histogram of the $j$-th dimension. 
Finally, we transform $T$ as $T_{\text{trans}} = T A_c A_r$.

\subsubsection{Gravity-based Movement}\label{ImprovedMovement}
UREM employs a gravity-based movement strategy to gather data points that share similar features or are frequently accessed together. As shown in Equation~\ref{gravity-equation}, each point \(a_i \in T_{\text{trans}}\) is assumed to be moved under the gravitational influence of other points within a predefined radius \(r\),  where closer points exert stronger attraction and induce larger displacements. Specifically, \(\vec{g_{iD}}\) and \(\vec{g_{iQ}}\) represent the movement from neighboring data points and query points, respectively. 
The relative influence of query points $Q$ on data points $D$ is controlled by a hyperparameter \(\tau\). Accordingly, \(a_i\) is updated to \(a_i'\) as follows:

\begin{equation}\label{gravity-equation}
\begin{aligned}
a_i' &= a_i + \big( \vec{g_{iD}} + \vec{g_{iQ}} \big) \\
\vec{g_{iD}} &= G \Bigg(
\sum_{j=1}^{k_i^{D}} \frac{\|d_{i1}-a_i\|^2}{\|d_{ij}-a_i\|^2}\,(d_{ij}-a_i)
\;+\;
\sum_{j=k_i^{D}+1}^{m_i^{D}} \frac{d_{ij}-a_i}{C}
\Bigg) \\
\vec{g_{iQ}} &= \tau\, G \Bigg(
\sum_{j=1}^{k_i^{Q}} \frac{\|q_{i1}-a_i\|^2}{\|q_{ij}-a_i\|^2}\,(q_{ij}-a_i)
\;+\;
\sum_{j=k_i^{Q}+1}^{m_i^{Q}} \frac{q_{ij}-a_i}{C}
\Bigg)
\end{aligned}
\end{equation}

In this formulation, the interaction radius \(r\) is typically set between \(5G\) and \(10G\), where \(G\) denotes the average nearest neighbor distance of each point in $T_\text{trans}$,
\(
G = \frac{1}{N} \sum_{i=1}^{N} \|a_{i1} - a_i\|,
\)
where \(a_{i1}\) is the nearest neighbor of \(a_i\) and \(N\) is the total number of points in \(T_{\text{trans}}\).
For a fixed \(a_i\), let \(\{d_{i1}, d_{i2}, \dots, d_{i m_i^{D}}\}\) denote all data neighbors from \(D_\text{trans}\) that lie within radius \(r\), ordered by increasing distance to \(a_i\); thus, \(m_i^{D}\) represents the total number of such data neighbors.
Similarly, let \(\{q_{i1}, q_{i2}, \dots, q_{i m_i^{Q}}\}\) denote the query point neighbors within \(r\), with \(m_i^{Q}\) representing the total number of query point neighbors.
The $k_i^{D}$-th data point is defined as  the furthest point of $a_i$ satisfying
\(\|d_{ij} - a_i\|^2 \le G \|d_{i1} - a_i\|\). Similarly, \(k_i^{Q}\) denotes the farthest query neighbor satisfying
\(\|q_{ij} - a_i\|^2 \le G \|q_{i1} - a_i\|\).
If no neighbor satisfies this condition, the corresponding \(k_i\) is set to zero,
and the first summation term in Equation~\ref{gravity-equation} is omitted.
$C$ is a constant slightly larger than 1, set as $1 + 10^{-1}$.
For the movement distance caused by the query points,
$\tau = \gamma/N_q$ means that more query points lead to smaller individual movements on other points, while fewer query points result in larger movements. 
Therefore, $\tau$ can prevent an excessive number of query points in $T_{\text{trans}}$ from affecting the outcome $T_{\text{adj}}$. Usually, $\gamma$ is set on a similar order of magnitude as the total number of data points.
This gravity-based movement not only encourages the gathering of data points that are similar in data characteristics or likely to be accessed together, but also eliminates ambiguous boundaries of data distribution.

\begin{figure}[h]
  \centering
\includegraphics[width=0.8\linewidth]{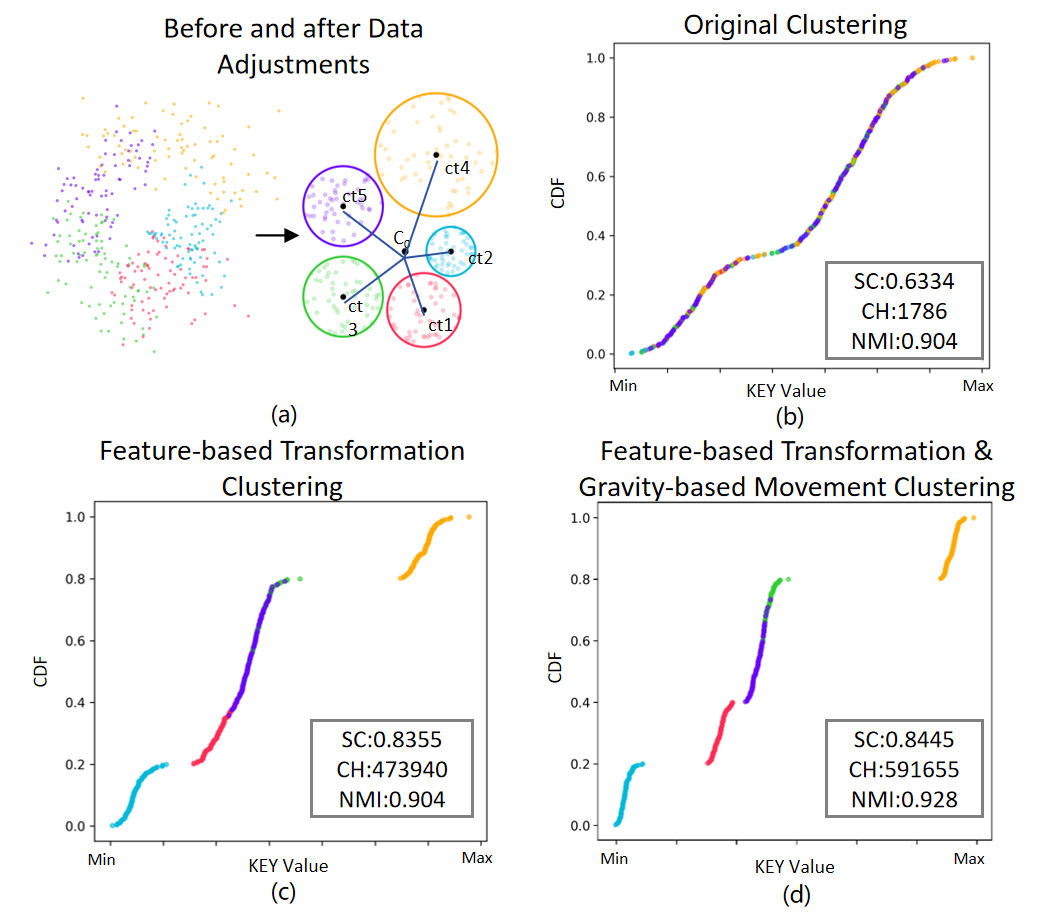}
  \caption{The CDF curves of KEY values and clustering evaluation results before and after data adjustments.}
  \label{offlinepartition}
\end{figure}

\subsection{Partitioning Process}\label{PartitioningProcess}
After applying data adjustments, UREM first performs a clustering. 
Since $T_{\text{adj}}$ includes both data points and historical query points, the query points must first be removed to obtain the pure data $D_{\text{adj}}$. 
We also compute the KEY value for each data point as $\text{KEY}(a_i) = ||a_i, ct_k|| + ||ct_k, ct_0||$, where $ct_k$ is the center of the cluster to which $a_i\in D_{\text{adj}}$ belongs, and $ct_0$ is the centroid of all cluster centers. As shown in Figure~\ref{offlineframework}, the data points with similar KEY values share similar characteristics and are likely accessed by similar queries.
This enables an effective one-dimensional mapping of $D_{\text{adj}}$. The primary objective of UREM is to enhance intra-cluster compactness while maximizing inter-cluster separability and to ensure the KEY values within each cluster are evenly distributed. Figure~\ref{offlinepartition} shows the effectiveness of UREM using three widely adopted metrics: Silhouette Coefficient (SC), Calinski–Harabasz Index (CH), and Normalized Mutual Information (NMI), and further illustrates the KEY value distribution through the Cumulative Distribution Function (CDF). As shown in Figure~\ref{offlinepartition}(a), the proposed data- and query-driven adjustment yields tighter clusters and fewer ambiguous boundary points.
Comparing Figures~\ref{offlinepartition}(b) and (c), feature-based transformation significantly improves SC, CH, and NMI scores, produces smoother CDF curves, and reduces KEY value overlap across clusters.
Similarly, Figures~\ref{offlinepartition}(c) and (d) indicate that gravity-based movement further enhances clustering quality.
Therefore, in the partitioning process, UREM partitions $D_{\text{adj}}$ by sorting them on KEY values and evenly dividing them into continuous ranges, then physically collocates the data to improve query performance.
We store the original data based on division results, which is $D_\text{opt}$. 

When handling queries, since the original data is stored, the results retrieved by the upper-level index built on $D_\text{opt}$ serve as the final outputs.

\section{Partial Layout Reorganization}\label{PartialOrganization}

\begin{figure}
  \centering
\includegraphics[width=0.9\linewidth]{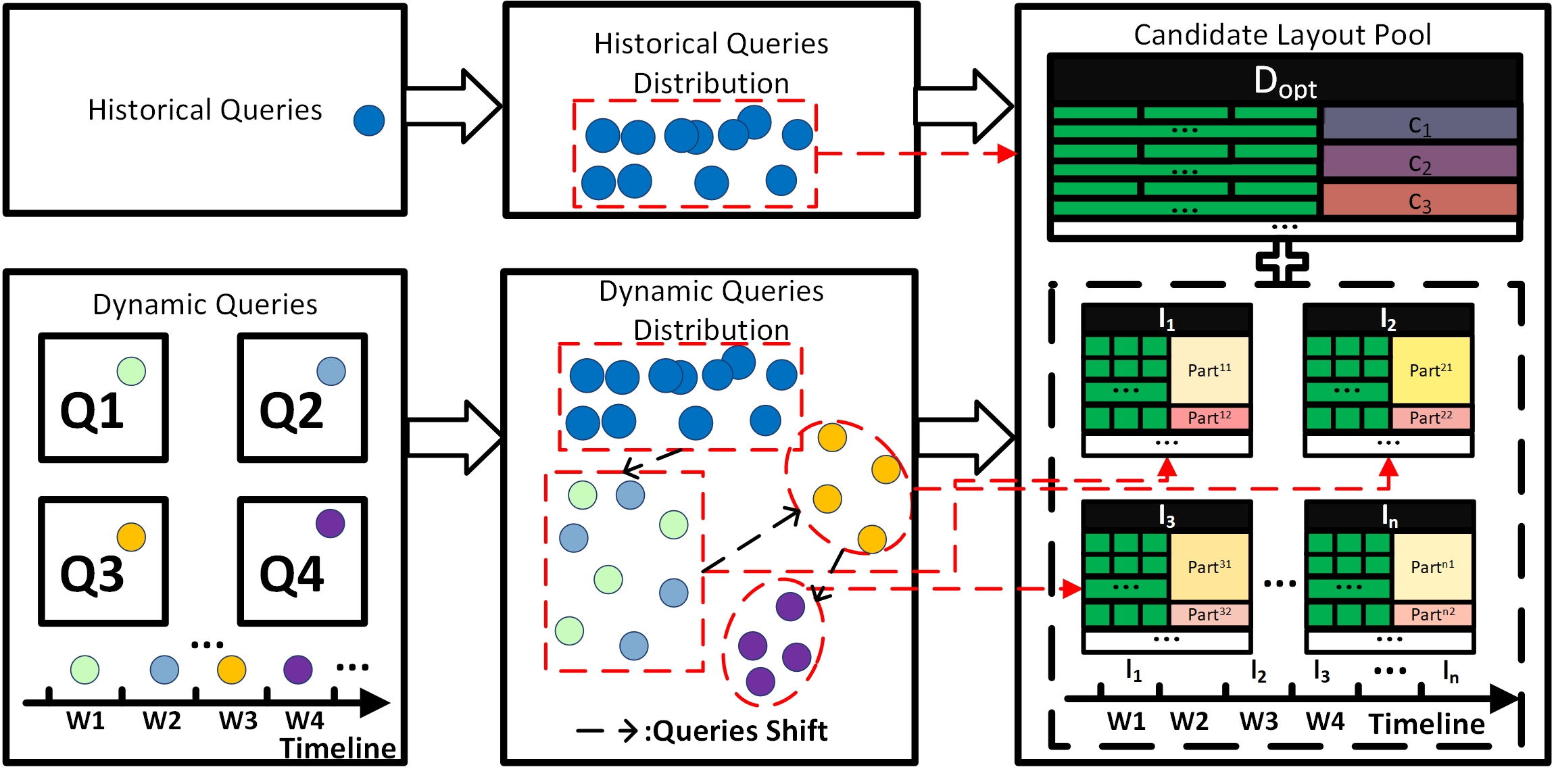}
  \caption{Partial layout reorganization.}
  \label{onlineframework}
\end{figure}

In real-world applications, queries are often dynamic and continuously shifting. 
To stabilize query performance, it is essential to adapt data layouts. 
An existing study has proposed a SOTA method, OREO~\cite {OREO}, which is adapted to shifting PBFs on multi-dimensional data. 
However, it cannot be applied to ANNS. 
To overcome this limitation, as shown in Figure~\ref{onlineframework}, UREM extends the OREO and supports flexible partial layout switching for ANNS in multi- or high-dimensional data under dynamic workloads.
Besides, UREM can further enhance the performance of OREO by leveraging $D_\text{opt}$ (described in Section~\ref{DeeplyDataOrganization}) during cold start and layout switching.

\subsection{Generating Partial Layouts}\label{GeneratingPartialReorganization}
To stabilize query performance under dynamic workloads, UREM adopts a sliding-window-based layout reorganization mechanism. 
Since the results of the queries within a sliding window typically cover only a small portion of the overall data, UREM reorganizes partial layouts instead of the entire dataset to reduce overhead.

\begin{algorithm}[!h]
\small    \caption{PartialLayoutGeneration}
    \label{alg:OrganizationGeneration}
    \renewcommand{\algorithmicrequire}{\textbf{Input:}}
    \renewcommand{\algorithmicensure}{\textbf{Output:}}

    \begin{algorithmic}[1]
        \Require data $D_\text{opt} \in \mathbb{R}^{N_d \times d}$, partition size $Size$, query points $Q_{W_i}$, ratio $\beta$
        \Ensure partial layout $l_i$

        \For{each $q_m \in Q_{W_i}$}
            \State $\rho(q_m) \gets \text{KDE}(q_m)$ \Comment{Kernel Density Estimation}
        \EndFor  

        \State $q_{r_i} \gets \arg\max \rho(q_m)$ \Comment{Select the representative query point}

        \For{each $a_m \in D_\text{opt}$}
            \State $\text{DIS}_i(a_m) \gets$ Euclidean distance $ ||a_m, q_{r_i}||$
        \EndFor

        \State $\texttt{Rank}(D_\text{opt}) \gets$ Sort the data in ascending order of $\text{DIS}_i(a_m)$
        \State $TotPart_i \gets$ The ceiling of $\beta \cdot N_d/ Size$ 
        \For{$b \gets1$ \textbf{to} $TotPart_i$}
            \State $\text{Part}^{ib} \gets$ Place the elements ranked within $[ (b{-}1) \cdot Size + 1 , \min(b \cdot Size, N_d) ]$ into the $b$-th partition \Comment{Evenly divide}
            \State $l_i \gets$ Add $\text{Part}^{ib}$ to $l_i$
        \EndFor

        \State \Return{$l_i$}
    \end{algorithmic}
\end{algorithm}

\begin{figure}
\centering
\includegraphics[width=0.9\linewidth]{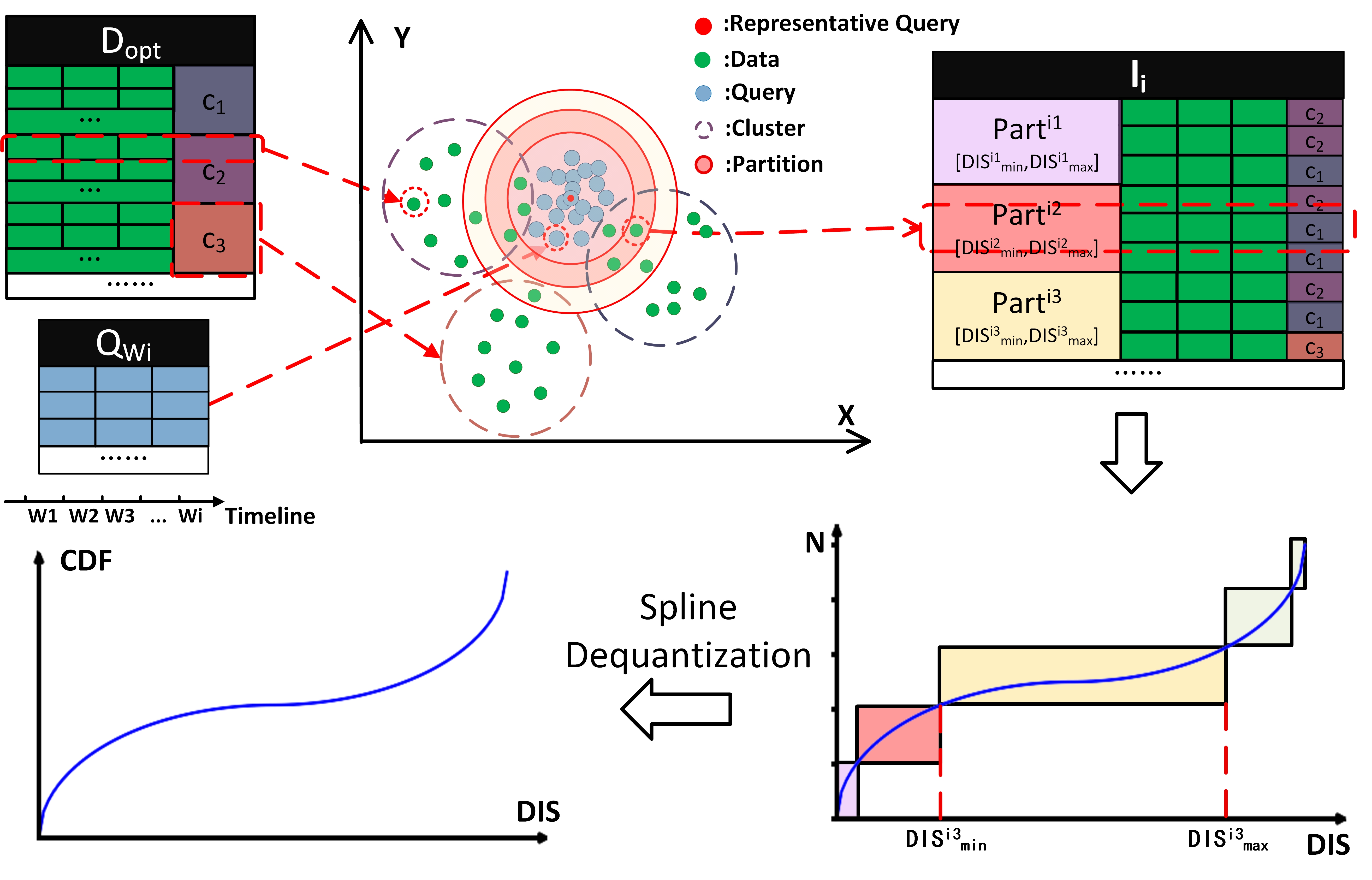}
  \caption{Partial layout generation and data distribution estimation.}
  \label{OnlinePartition}
\end{figure}

Algorithm~\ref{alg:OrganizationGeneration} demonstrates how UREM constructs a new partial layout $l_i$ based on query points $Q_{W_i}$ within a sliding window $W_i$. 
As shown in Figure~\ref{OnlinePartition}, since ANNS measures similarity by the distance between a query point $q$ and data points, UREM selects a representative query point $q_{r_i}$ for each $W_i$ and groups data points with similar distances to $q_{r_i}$ into the same partition.
Kernel Density Estimation is used to estimate the density of query points within $W_i$, and the point with the highest density is chosen as $q_{r_i}$ (Lines 1–3).
Then, UREM computes the distance $\text{DIS}_i(a_m)$ between each data point $a_m$ and $q_{r_i}$, followed by sorting them (Lines 4–6). 
This sorting helps group data points that may belong to separate clusters under static workloads into the same partition when queries shift, thereby reducing cross-partition access rates. 
UREM then divides the sorted data points evenly (Lines 8-10) to get a newly generated layout $l_i$. 
For each $l_i$, UREM records the DIS range between all points in the $j$-th partition and $q_{r_i}$ as the partition's metadata $\left[ \text{DIS}_\text{min}^{ij}, \text{DIS}_\text{max}^{ij} \right]$, where $\text{DIS}_\text{min}^{ij}$ denotes the minimum distance from $q_{r_i}$ to any point in the $j$-th partition, and $\text{DIS}_\text{max}^{ij}$ denotes the maximum distance.
Since queries within $Q_{W_i}$ typically access only a small subset of $D_\text{opt}$, UREM selectively constructs a partial layout $l_i$ associated with $Q_{W_i}$ through the parameter $\beta$, $0 < \beta \le 1$ (Line 7), reducing reorganization costs while ensuring that the new $l_i$ covers the majority of query results within the $W_i$. 

\begin{theorem}\label{thm1}
The total cost under dynamic workloads is minimized when the size $N_l$ of the partial layout $l_i$ satisfies the condition $ \frac{k(1-p)^{k-1}}{N_d} \left(\log N_l - C \right) + \frac{1-(1-p)^k}{N_l} = - \frac{\lambda \cdot B}{A}$, where $p=\frac{N_l}{N_d}$, $A$ is the total number of queries, $B$ is the number of partial layouts in the candidate layout pool, $C$ refers to the query cost of the upper-layer index on the data $D_\text{opt}$, $\lambda$ is the normalization coefficient, and $k$ represents the number of target neighbors in ANNS queries.
\end{theorem}

\begin{IEEEproof}
Under dynamic workloads, the total cost consists of both the query cost and the storage overhead of partial organizations, which is defined as
$\text{TotalCost} = \text{QueryCost} + \text{StorageCost}$, where the \text{QueryCost} is $A \left[ (1-(1-p)^{k}) \log N_l + (1-p)^{k}C \right]$ (discussed in detail in Section~\ref{discussion}), and the \text{StorageCost} is the cost of storing $B$ partial layouts, i.e., $B \cdot N_l$. To place $\text{QueryCost}$ and $\text{StorageCost}$ on a unified scale, a normalization coefficient $\lambda$ is multiplied to $\text{StorageCost}$, i.e., $\lambda \cdot B \cdot N_l$.
To minimize the total cost, we take the derivative of the cost function and set it to zero, which is $ \frac{k(1-p)^{k-1}}{N_d} \left(\log N_l - C \right) + \frac{1-(1-p)^k}{N_l} = - \frac{\lambda \cdot B}{A}$. 
\end{IEEEproof}

\subsection{Expanding Candidate Layout Pool}\label{ExpandingPool}
Retaining all newly generated partial layouts leads to high storage overhead, and similar queries can share layouts to reduce this cost. 
Therefore, UREM introduces a candidate layout pool $L$ containing the optimized layout $D_\text{opt}$ and partial layouts generated from sliding windows.
During the cold-start phase, $D_\text{opt}$ serves as the initial layout in $L$. As queries shift over time, new partial layouts are generated, adding to $L$ and used for queries. 

UREM adopts the Reservoir-Time Biased Sampling (R-TBS)~\cite{R-TBS, OREO} algorithm to draw some query samples of size $S$ from queries $Q_{W_i}$ within the sliding window $W_i$. 
For each $l_i$, the query samples are estimated to produce a cost vector $C_i = (c(q_1, l_i), \ldots, c(q_S, l_i))$, where $c(q_j, l_i)$ represents the estimated query cost of executing the $j$-th query sample on layout $l_i$ (the estimation process is introduced in Section~\ref{FindingOptimal}). 
The difference between layouts $l_i$ and $l_j$ is defined as the distance between their cost vectors $\text{Diff}(C_i, C_j)=||C_i, C_j||$. 
UREM defines a distance threshold $\varepsilon \in [0, 1]$. 
A new partial layout $l_i$ is added to $L$ after $W_i$ finishes only if its $C_i$ differs from those of all existing layouts in the pool $L$ by more than $\varepsilon$. 
Therefore, as the pool grows, it becomes increasingly difficult for new partial layouts to be accepted, and the number of layouts in the pool $L$ is finite.

\begin{theorem}
The cost vector $C_i \in \mathbb{R}^m$ represents the cost of $m$ queries within a sliding window on $l_i$. 
For any layout $l_i$ and $l_j$ ($i\ne j$) in pool,  $C_i,C_j \in [0, 1]^m$ and $\text{Diff}(C_i, C_j) > \varepsilon$.
Then, the pool size $n$ is bounded as $O\left( \left( 1/\varepsilon+1 \right)^m \right)$.
\end{theorem}

\begin{IEEEproof}
The cost vector $C_i$ is an $m$-dimensional unit hypercube $[0,1]^m$. 
The constraint $\text{Diff}(C_i, C_j)> \varepsilon$ implies that each $C_i$ must be at least $\varepsilon$ apart from all others, which corresponds to placing non-overlapping spheres of radius $\varepsilon/2$ centered at each $C_i$ within the unit hypercube. 
The total number of such balls whose centers lie within $[0,1]^m$ is bounded by $\frac{V_m((1+\varepsilon)/2)}{V_m(\varepsilon/2)} = O\left( \left(1/\varepsilon+1\right)^m \right)$, where $V_m(r)$ denotes the volume of an $m$-dimensional ball of radius $r$.
\end{IEEEproof}
To further control the storage overhead of $L$, UREM can periodically remove $l_j \in L$ that yields similar query cost to those of other existing ones.

\begin{algorithm}[h!]
\small    \caption{EstimatedQueryCost}
    \label{alg:EstimatedQueryCost}
    \renewcommand{\algorithmicrequire}{\textbf{Input:}}
    \renewcommand{\algorithmicensure}{\textbf{Output:}}

    \begin{algorithmic}[1]
        \Require partial layout $l_i \in L$, query point $q$, representative query point $q_{r_i}$, the total number of partitions within partial layout $TotPart_i$
        \Ensure estimated query cost $c(q,l_i)$

        \State Construct the partition distribution histogram $h_i$ of $l_i$
        \State $F_{l_i}(q_{r_i},x) \gets \textsc{SplineDeq}(h_i)$ \Comment{Spline dequantization}
        
        \State $r\gets$ Compute by using the inverse of $F_{l_i}$ \Comment{Estimate radius}

        \State $P\gets$ The distance  between $q$ and $q_{r_i}$, which is $\text{DIS}_i(q)$
        \State $AcsPart_i\gets \sum_{j} \mathbf{1} \left\{ \left[ \text{DIS}_\text{min}^{ij}, \text{DIS}_\text{max}^{ij} \right] \cap \left[ P - r, P + r \right] \neq \emptyset \right\}$

        \State $c(q,l_i)\gets AcsPart_i/TotPart_i$ \Comment{Estimate query cost}

        \If{$c(q,l_i)=0$}
        \State $c(q,l_i)\gets1$ \Comment{Full scan of the dataset}
        \EndIf

        \State \Return{$c(q,l_i)$}
    \end{algorithmic}
\end{algorithm}

\subsection{Finding Optimal Partial Layout}\label{FindingOptimal}

To stabilize query performance, UREM estimates the accumulated query cost of each layout and predefines a threshold to determine when and which partial layout to switch to.  

Algorithm~\ref{alg:EstimatedQueryCost} illustrates how UREM estimates the query cost of executing a query on partial layout $l_i$.
Since $l_i$ records the DIS range as metadata of each partition, it is necessary to convert the $k$-ANNS query, which finds the $k$-th nearest neighbours of query point $q$, into a distance-based query $(q,r)$, where $r$ is the query radius, enabling query cost estimation. 
Prior research~\cite{PM-LSH} has shown that, given the distance distribution $F(a_1,x) = P[||a_1, a_2|| \leq x]$ over a dataset $D\in \mathbb{R}^{N}$, it is possible to determine a query radius $r$ such that $N \cdot F(q,r) = k + C$, implying that at least $k$ nearest neighbor points are expected to lie within the hypersphere centered at $q$ with radius $r$, where $C$ is a constant. 
So, we need to obtain a distance distribution $F_{l_i}(q_{r_i},x)$ of partial layout $l_i$ generated in $W_i$ based on its representative query point $q_{r_i}$ for query cost estimation, as shown in Figure~\ref{OnlinePartition}. We first construct a histogram, which shows the data distribution of partitions within $l_i$ (Line 1). 
Then, we apply spline dequantization~\cite{FACE} to recover a smooth approximation of the distribution $F_{l_i}(q_{r_i},x)=P[\text{DIS}_i(a) \leq x]$ (Line 2) and get the estimated query radius $r$ (Line 3).
By leveraging the metadata of each partition and $r$, we can estimate the number of partitions potentially accessed by a query, denoted by $AcsPart_i$ (Lines 4-5). 
Therefore, the estimated query cost $c(q,l_i)$ equals $AcsPart_i$ divided by the total number of partitions $TotPart_i$(Line 6). 
Since $c(q, l_i) = 0$ means the query result is not contained in $l_i$, a full data scan is required. Therefore, we set $c(q, l_i) = 1$ in this case (Lines 7–8).

During the continuous shifting of queries, we record the cumulative query cost of $l_i$ as $ctotal_i=\sum c(q, l_i)$.
When $ctotal_i$ exceeds a predefined threshold $\alpha$ ~\cite{OREO}, it indicates that the current layout $l_i$ has become suboptimal and triggers an optimal layout switching. 
To select a new optimal layout, UREM assigns each $l_j$ a weight $w_j=\mathop{\max}(0,(\alpha-ctotal_j))/\alpha$, which is proportional to the average number of irrelevant data points skipped by $l_j$ during past queries. When a switching is triggered, UREM selects the next $l_j$ with probability $p_j=\frac{w_j }{\sum_{j=1,\dots,n} w_j}$, and designates it as the new optimal partial layout. 
After switching, UREM resets $ctotal_i$ of $l_i\in L$, enabling the system to adapt effectively to future queries.

\begin{algorithm}[t!]
\small
\caption{OnlineQuery}
\label{alg:onlinequery}
\renewcommand{\algorithmicrequire}{\textbf{Input:}}
\renewcommand{\algorithmicensure}{\textbf{Output:}}
\begin{algorithmic}[1]
\Require Data $D_\text{opt} \in \mathbb{R}^{N_d \times d}$, query point $q$, optimal partial layout $l_{opt}$, number of neighbors $k$, and parameter $a$ 
\Ensure $k$ nearest neighbors of query point $q$

\State $can \gets \text{Initialize a candidate results max-heap}$

\State $i \gets \textsc{BTreeSearch}(l_{opt}.\texttt{BTree},  \text{DIS}_{opt}(q))$

\State $left \gets i - 1$, $right \gets i + 1$ \Comment{Bidirectional neighbor search}
\State  Add $l_{opt}[i]$ to $can$
\State $ \texttt{max} \gets \text{Initialize the maximum distance as } \text{DIS}_{opt}(l_{opt}[i])$  
\While{$left \geq 0$}
    \If{$can.Size<ak$ \textbf{or} $\text{DIS}_{opt}(l_{opt}[left]) < \texttt{max}$}
        \State Add $l_{opt}[left]$ to $can$ and assign $\texttt{max}$ as $\text{DIS}_{opt}(can[0])$
         \EndIf
         \If{$|\text{DIS}_{opt}(l_{opt}[left])-\text{DIS}_{opt}(q)| \geq \texttt{max}$}
     \State \textbf{break}
     \EndIf
    \State $left \gets \text{Move the } left \text{ pointer one position to the left}$
    \If{$left < 0$} \Comment{Touch boundary}
    \State Access the boundary of $l_{opt}$\EndIf
\EndWhile
\While{$right < l_{opt}.Size$}     
     \State The $right$ pointer repeats the operations in Lines 7–10.
    \State $right \gets \text{Move the } right \text{ pointer one position to the right}$ 
    \If{$right \geq l_{opt}.Size$} \Comment{Touch boundary} 
    \State Access the boundary of $l_{opt}$\EndIf
\EndWhile

\If{$can.Size < k$ \textbf{or} previously accessed the boundary}
    \State $res \gets \texttt{Index}(D_\text{opt},q, k)$ \Comment{Fallback to full index scan}\EndIf
\State \textbf{else} $res \gets can.\texttt{Topk}$

\State \Return $res$

\end{algorithmic}
\end{algorithm}

\subsection{Query Process}\label{OnQueryProcess}
During the query process, UREM first attempts to query on the optimal partial layout $l_{opt}$. 
If $l_{opt}$ does not contain all query results, the search falls back to the full dataset $D_\text{opt}$ using the upper-level index (Section~\ref{DeeplyDataOrganization}). 
As illustrated in Algorithm~\ref{alg:onlinequery}, UREM maintains a candidate result set $can$, which is ordered by the DIS values of data to record the potential query results (Line 1). 
During query execution, UREM first uses a B-tree to locate the position of the data point in $l_{opt}$ closest to $q$ and uses two pointers to operate a bidirectional neighbor search (Lines 2-6, 14). 
Since the data is sorted based on DIS values in $l_{opt}$, such ordering does not strictly reflect proximity in the original hyperspace. 
Therefore, UREM incrementally explores data points and adds them to $can$ if $can$ is not full or the distance between the data point and query point $q$ is smaller than the largest one in $can$ (Lines 7–8). 
The search continues until $|\text{DIS}(L_{opt}[pointer])-\text{DIS}(q)| \geq \texttt{max}$ (Lines 9–10), since $|\text{DIS}(L_{opt}[pointer])-\text{DIS}(q)| \leq ||L_{opt}[pointer], q||$. 
However, if the query assesses the boundary of $l_{opt}$, it suggests that $l_{opt}$ may not contain all $k$ nearest neighbors (Lines 12-13, 17-18). 
Therefore, UREM needs to perform a fallback search on the full dataset $D_\text{opt}$, along with its associated index structures (Lines 19-20).

\section{Experiment}~\label{experiments}
\subsection{Experimental Setup}
\subsubsection{Environment}
We implement the experiments on several platforms, including a relational database, a vector database, and a data lake. 
The relational database and vector database are deployed on a single-node server, whose hardware configuration includes 64-bit Ubuntu 20.04, two Intel(R) Xeon(R) Gold 6226 CPUs @ 2.90 GHz, 256 GB RAM.
The data lake is deployed in a distributed cluster environment consisting of three nodes, where each node is equipped with 64-bit Ubuntu 18.04, Intel(R) Core(T) i7-11700F CPU @ 2.50 GHz, and 16 GB RAM.

\subsubsection{Datasets and Workloads}

\begin{table}[!h]
  \caption{Summary of datasets.}
  \label{dataset}
  \begin{tabularx}{0.45\textwidth}{m{2cm}<{\centering}m{1.25cm}<{\centering}m{0.7cm}<{\centering}m{1.3cm}<{\centering}m{1cm}<{\centering}}
    \toprule
    Datasets & Type & Dim. & Cardinality & Size(GB) \\
    \midrule
    Stocks~\cite{Stocks} & real & 4 & 20M & 1.3 \\
    Taxi~\cite{Taxi} & real & 5 & 184M & 11.8 \\
    OSM~\cite{osm} & real & 6 & 105M & 5.04 \\
    Dominick~\cite{Dominick} & real & 7 & 38M & 1.6 \\
    TPC-H~\cite{TPC-H} & real & 7 & 300M & 16.8 \\
    Forest~\cite{Forest} & real & 12 & 0.56M & 1.5 \\
    Color~\cite{Color} & real & 32 & 1.28M & 4.2 \\
    Yelp~\cite{Yelp} & real & 64 & 6.99M & 13.2 \\
    SIFT1B~\cite{SIFT} & real & 128 & 1B & 16.2 \\
    DEEP1M~\cite{DEEP} & real & 256 & 1M & 4.5 \\
    GIST1M~\cite{GIST} & real & 960 & 1M & 12.1 \\
    GuassMix & synthetic & 3-960 & 1-100M & N.A. \\
    \bottomrule
  \end{tabularx}
\end{table}

Table~\ref{dataset} shows the eleven real-world datasets and one synthetic dataset we used in our experiments. 
Specifically, the synthetic dataset GaussMix, whose dimensionality and cardinality can be manually controlled, is utilized to assess the scalability of the indexes.

Under static workloads, if the dataset includes queries, we use it directly. 
Otherwise, we generate synthetic queries. 
Each query consists of a data point, which is randomly sampled from the dataset, and an associated parameter~\cite{mqrld}. 
For PBF, the parameter is query selectivity, representing the percentage of data expected to be retrieved. 
We default it as 10\%. 
For $k$-ANNS, the parameter is the number of nearest neighbors $k$, set to 100 by default. 
Under dynamic workloads, to better simulate the continuously shifting queries in real-world applications, we divide the queries into 15 biased sliding windows, each containing 2,000 queries~\cite{querywindow}.

\subsubsection{Metrics}
For PBF, we use query time to evaluate performance~\cite{Qd-tree, MTO}. 
For $k$-ANNS, we adopt query time and recall~\cite{spann, DET-LSH} as evaluation metrics. 
Accordingly, we use the following two key metrics:

\noindent(1) Avg Query Time: To accurately evaluate PBF query performance, each experiment is repeated five times, and the average query time is reported. We also assess the ANNS query performance of multi-dimensional indexes by average query time at a recall of 1.

\noindent(2) Query Time-Recall: For ANNS queries on high-dimensional indexes, we record recall across different query times. The recall is defined as the proportion of true nearest neighbors retrieved by the $k$-ANNS out of all actual relevant neighbors, reflecting the completeness of the search.

\subsubsection{Baselines}
We compare the query performance of the original indexes and UREM-enhanced indexes (i.e., UREM + index), as listed in Table~\ref{competitor}, under both static and dynamic workloads. 

\begin{table}[t!]
  \caption{The multi- and high-dimensional indexes used as baselines in the experiments.}
  \label{competitor}
  \begin{tabularx}{0.47\textwidth}{m{0.7cm}<{\centering}m{1cm}<{\centering}m{5.5cm}<{\raggedright}}
\toprule
Query & Dim. & Indexes \\
\midrule
PBF & multi- & R*-tree~\cite{r_tree2}, ML~\cite{Ml-index}, LISA~\cite{LISA}, Flood~\cite{Flood}, ZM~\cite{ZM}, Qd-tree~\cite{Qd-tree}, Tsunami~\cite{Tsunami}, MQRLD~\cite{mqrld}\\ \hline
ANNS & multi- & M-tree~\cite{m_tree}, R*-tree~\cite{r_tree2}, SPB-tree~\cite{spb_tree}, ML~\cite{Ml-index}, LISA~\cite{LISA}, LIMS~\cite{LIMS}, MQRLD~\cite{mqrld}\\ \hline
ANNS & high- & FLEX~\cite{FLEX}, LIRA~\cite{LIRA}, DB-LSH~\cite{DB_LSH}, DET-LSH~\cite{DET-LSH}, IVFADC~\cite{IVFADC}, RaBitQ~\cite{rabitQ}, HNSW~\cite{HNSW}, LSH-APG~\cite{LSH-APG}, HAMG~\cite{hamg}\\
\bottomrule
\end{tabularx}
\end{table}

\begin{figure*}[b]
  \centering
  \begin{minipage}[t]{\linewidth}
  \centering
\includegraphics[width=0.9\linewidth]{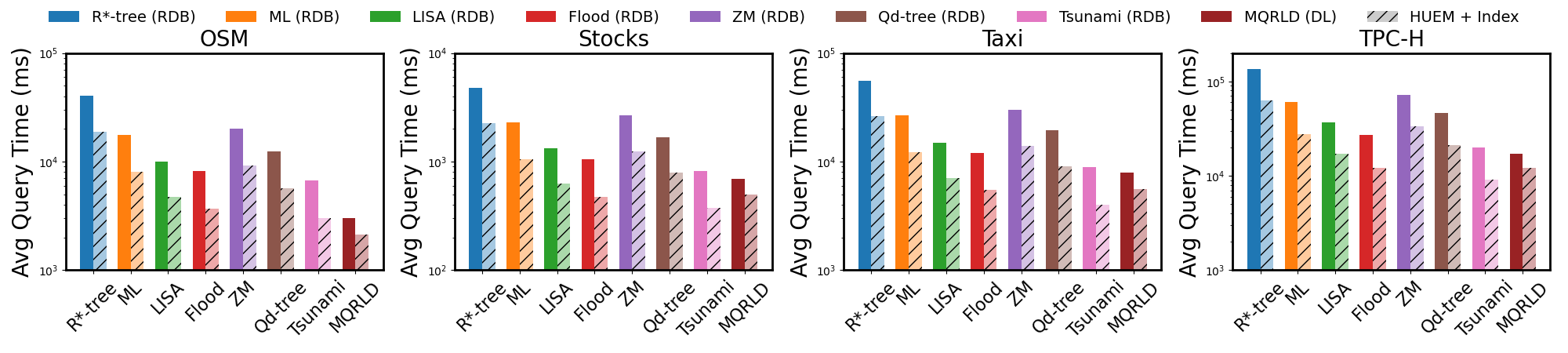}
  \caption{The performance comparison of PBF queries on multi-dimensional indexes before and after UREM enhancement.}
  \label{offlinePBF}
  \end{minipage}
  \begin{minipage}[t]{\linewidth}
  \centering
\includegraphics[width=0.9\linewidth]{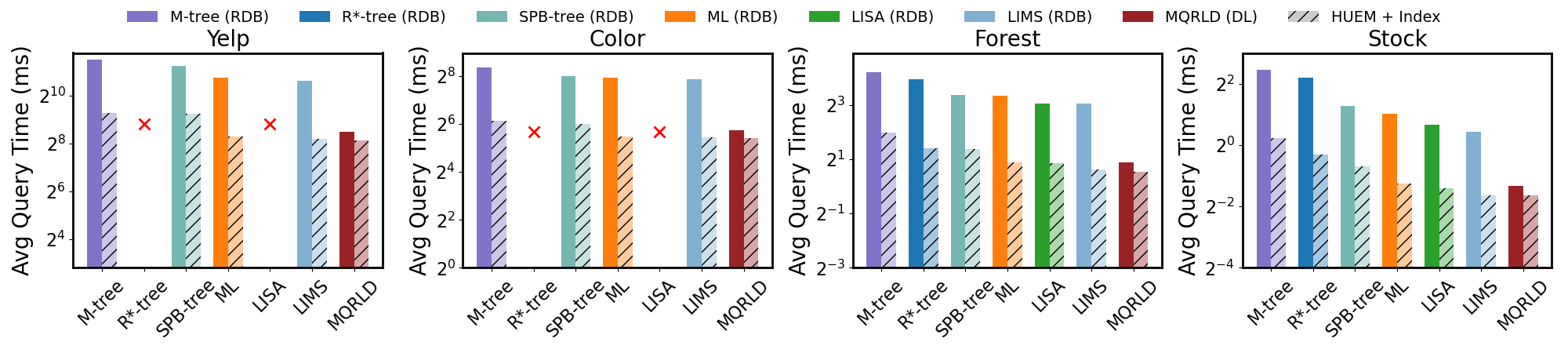}
  \caption{The performance comparison of ANNS queries on multi-dimensional indexes before and after UREM enhancement.}
  \label{offline_multianns}
  \end{minipage}
\end{figure*}

\subsubsection{Parameter Settings and Implementation}
Under static workloads, we set the parameter $\gamma$ as $N_d$, where $N_d$ denotes the number of data points. 
Under dynamic workloads, we use a dataset of size $1$M and perform $50$-NN queries, with $\gamma$ set to 0.1~\cite{alex}. Under these settings, the size $N_l$ of each partial layout is set to $1/50$ of the original dataset size $N_d$, that is, $\beta = 1/50$.
Additionally, following existing research~\cite{OREO}, the layout switching threshold $\alpha$ is set to 60 to determine when a layout switching should be triggered, and the distance threshold $\varepsilon$ is set to 0.1 to decide whether a newly generated layout should be included in the layout pool.

\subsection{Evaluation of UREM under Static Workloads}\label{experimentsStatic}
\subsubsection{Evaluation of PBF for multi-dimensional indexes}\label{multi-PBF}

Figure~\ref{offlinePBF} shows that UREM effectively enhances the PBF query performance of multi-dimensional indexes across four datasets: OSM, Stocks, Taxi, and TPC-H.
The experimental results indicate that UREM improves the efficiency of the traditional multi-dimensional index R*-tree by up to 2.14$\times$. 
This is because their query performance relies on reliable data partitioning during index construction, and UREM provides an effective data partitioning strategy that boosts their query performance. 
For learned indexes such as ML, LISA, Flood, ZM, Qd-Tree, and Tsunami, UREM achieves maximum speedups of 2.14$\times$-2.23$\times$. Even for MQRLD, which already employs its own layout optimization, UREM still delivers a 1.42$\times$ improvement. Notably, the UREM-enhanced R*-Tree even surpasses recent advanced indexes such as ML and ZM on certain datasets, highlighting UREM’s strong potential to elevate traditional indexes widely used in practice.

\subsubsection{Evaluation of ANNS for multi-dimensional indexes}\label{multi-ANNS}

As shown in Figure \ref{offline_multianns}, UREM significantly improves the query efficiency of $k$-ANNS on multi-dimensional indexes, where $k=100$. 
We evaluate its performance on four datasets: Yelp, Color, Forest, and a synthetic GaussMix dataset (cardinality = 1M, dimension = 6). 
Similar to the results observed for PBF, when the recall is 1, UREM improves the query time of tree-based indexes M-tree and SPB-tree by up to 4.71$\times$ and 3.98$\times$, respectively, thanks to its effective layout optimization. 
For learned indexes ML, LIMS, and MQRLD, UREM achieves up to 5.54$\times$, 5.36$\times$, and 1.28$\times$ improvements. 
We exclude R*-tree and LISA results on the Color and Yelp datasets due to their exponential query cost growth under high dimensionality.
For low-dimensional datasets such as Forest and GaussMix, UREM improves R*-tree and LISA performance by up to 5.73× and 4.55×, respectively.
Similarly, for multi-dimensional ANNS queries, UREM-enhanced R*-tree consistently outperforms recent advanced indexes, highlighting the effectiveness of UREM.

\subsubsection{Evaluation of ANNS for high-dimensional indexes}\label{high-ANNS}

\begin{figure*}[!t]
  \centering
\includegraphics[width=0.9\linewidth]{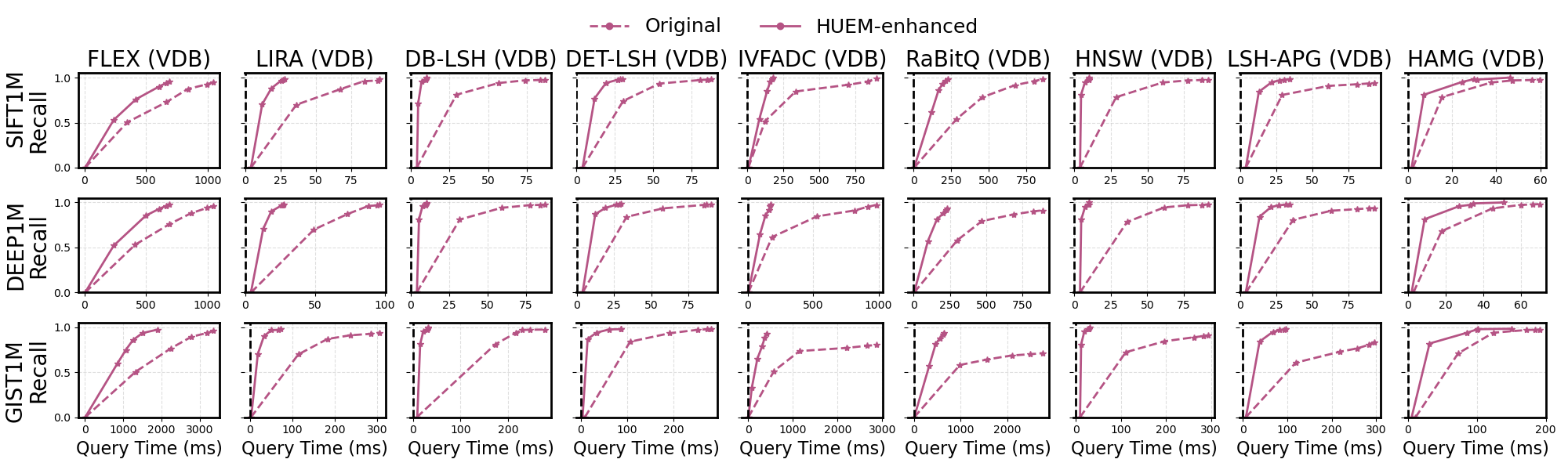}
  \caption{The performance comparison of ANNS queries on high-dimensional indexes before and after UREM enhancement.}
  \label{offline_highanns}
\end{figure*}

Figure~\ref{offline_highanns} demonstrates that UREM effectively enhances the $k$-ANNS query performance of high-dimensional indexes. 
We evaluate nine indexes on three datasets: SIFT1M, DEEP1M, and GIST1M. 
The results indicate that as the data dimensionality increases, the angle between the time-recall curves of the original indexes and those of the UREM-enhanced indexes becomes larger, suggesting a stronger enhancement effect. 
UREM notably boosts the performance of HNSW and DB-LSH, achieving speedups of up to 9.18$\times$ and 8.63$\times$, respectively, when recall equals 0.98. 
This improvement stems from UREM’s ability to reduce boundary data points, leading to more reliable index construction.
For other indexes, UREM also delivers 2$\times$ to 6$\times$ query performance improvements. 
Although the gains for these indexes are smaller than HNSW and DB-LSH, such improvements can still greatly reduce query costs in real-world applications.

\subsubsection{Scalability}\label{scalability}
\begin{figure}[t!]
  \centering
\includegraphics[width=0.9\linewidth]{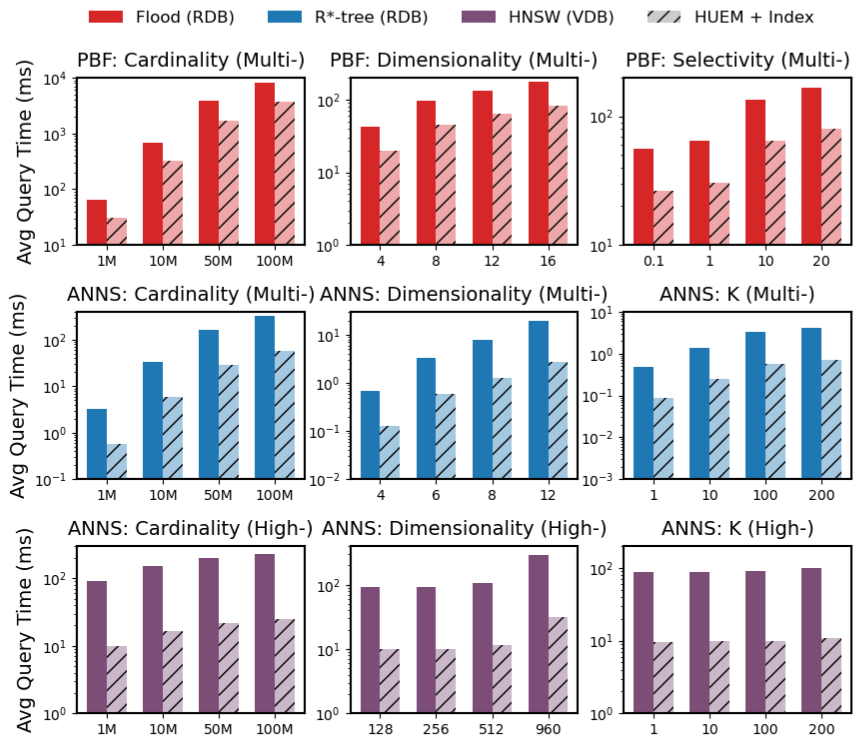}
  \caption{UREM scalability evaluation.}
  \label{offline_scalability}
\end{figure}

To investigate the scalability of UREM, we use the synthetic dataset GaussMix. 
We evaluate index performance before and after UREM enhancement using datasets of varying sizes and dimensions under default settings (cardinality: 1M; dimensionality: 6 for multi-dimensional and 128 for high-dimensional data; PBF selectivity: 10\%; $k$-ANNS: 100). 
As shown in Figure~\ref{offline_scalability}, for PBF, we select Flood to evaluate query performance enhancement, whereas for ANNS, we use the multi-dimensional index R*-tree and the high-dimensional index HNSW to assess UREM’s performance. 
The experimental results consistently demonstrate that UREM effectively improves query performance across different data cardinalities, dimensionalities, and query parameter configurations. 
Furthermore, the enhancement effect becomes stronger as the data cardinality increases, the dimensionality grows higher, and the query becomes more complex.

\begin{figure}[t!]
  \centering
\includegraphics[width=0.9\linewidth]{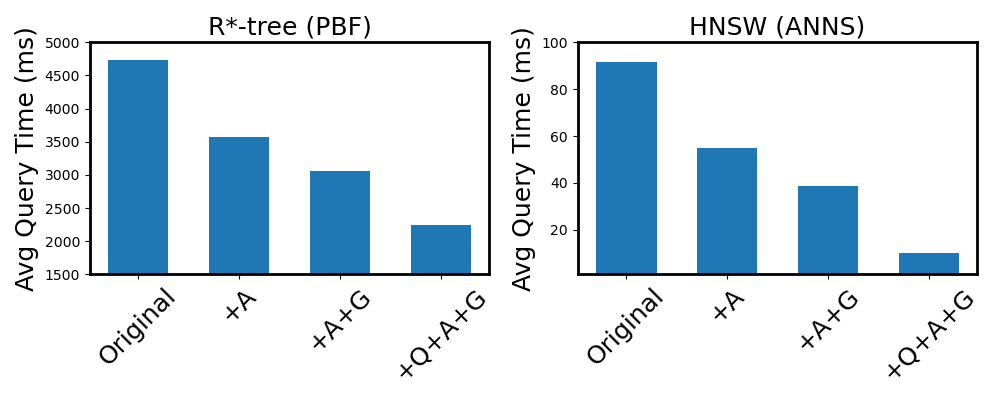}
  \caption{Ablation experiments of UREM.}
  \label{offline_ablation}
\end{figure}

\begin{figure*}[h!]
  \centering
\includegraphics[width=0.9\linewidth]{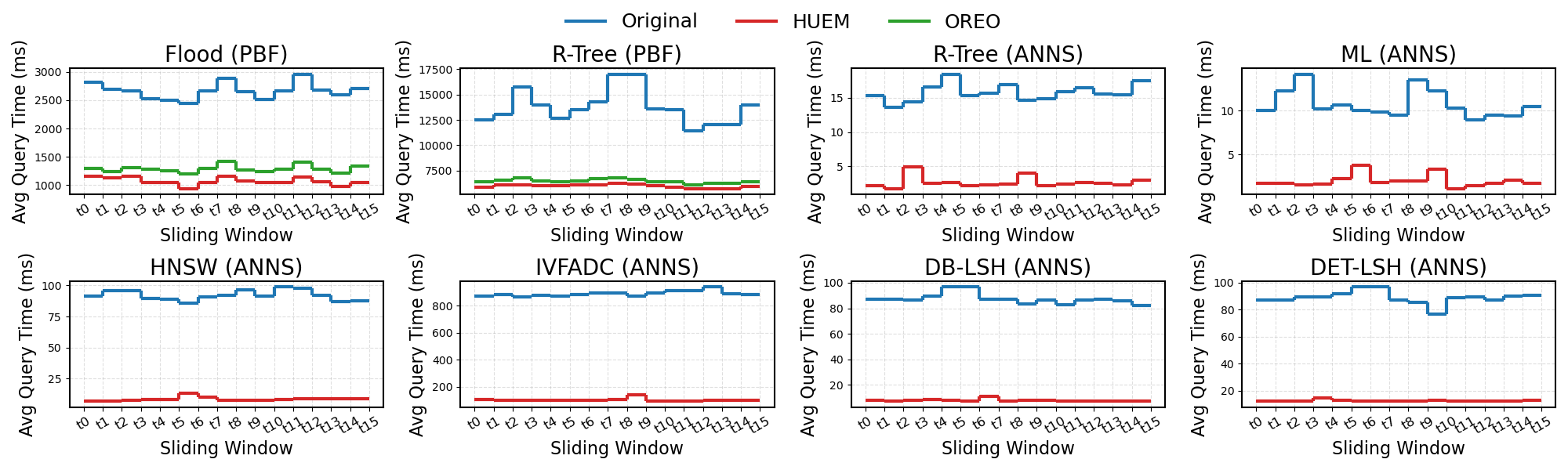}
  \caption{The performance comparison of different indexes in dynamic workloads before and after UREM enhancement.}
  \label{online_com}
\end{figure*}

\subsubsection{Ablation Experiments}\label{AB}
We conduct ablation experiments using R*-tree and HNSW to evaluate the effectiveness of UREM on multi- and high-dimensional indexes, as shown in Figure~\ref{offline_ablation}.  
In this context, “+A” denotes the application of feature-based transformation to the data (Section~\ref{ImprovedTransformation}), “+G” refers to the application of gravity-based movement (Section~\ref{ImprovedMovement}), and “+Q” indicates the application of vertical concatenation of the data points with query points (Section~\ref{VerticallyConcatenate}). 
The experimental results show that each step in UREM under static workloads can effectively improve the query performance. Comparing “original” with “+A” demonstrates that the feature-based transformation significantly enhances query performance, while the comparison between “+A” and “+A+G” highlights the effectiveness of the gravity-based movement. In addition, applying “+Q” also leads to a notable performance improvement.

\subsection{Evaluation of UREM under Dynamic Workloads}

\begin{figure}[t]
  \centering
\includegraphics[width=\linewidth]{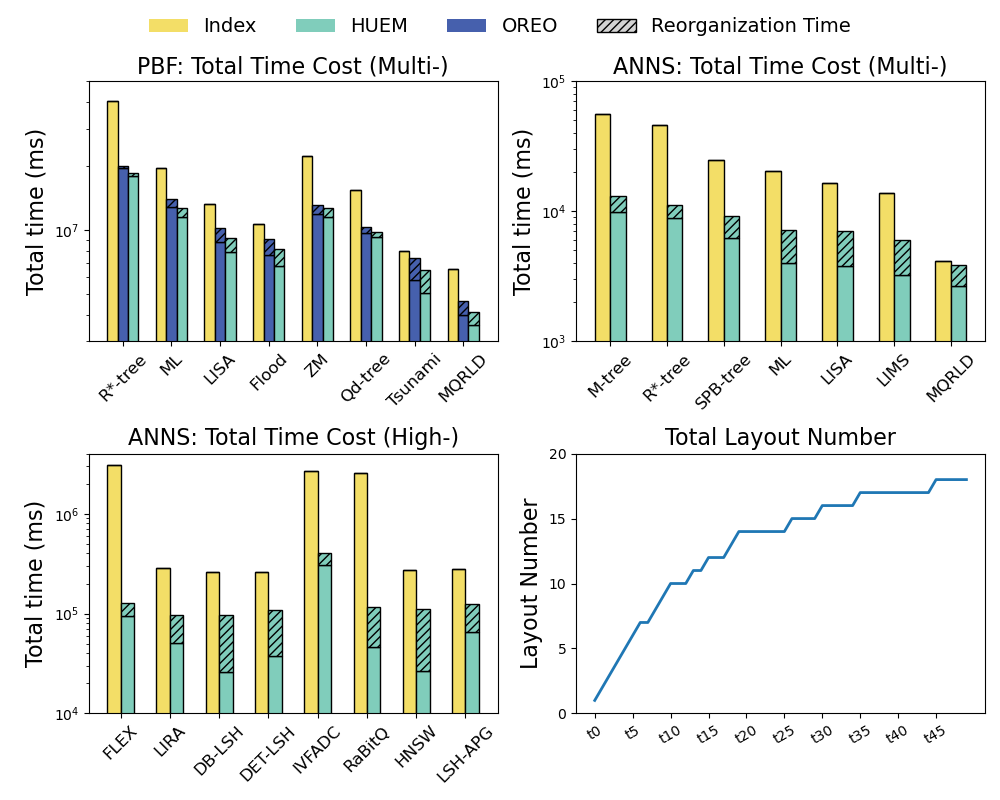}
\caption{(a)-(c) Comparison of the total time cost of indexes before and after UREM  enhancement under dynamic workloads, and (d) the analysis of the number of layouts in the candidate layout pool.}
  \label{online_layouts}
\end{figure}
\subsubsection{Evaluation of Query Performance when Queries Shift}\label{dynamic}
We validate the effectiveness of UREM under dynamic workloads. Due to space limitations, we present results on two indexes supporting PBF over multi-dimensional data (Flood and R*-tree), two indexes supporting ANNS over multi-dimensional data (R*-tree and ML), and four indexes supporting ANNS over high-dimensional data (HNSW, IVFADC, DB-LSH, and DET-LSH), demonstrating their performance improvements under dynamic workloads.
We use the Dominick dataset for PBF as multi-dimensional data, the Forest dataset for ANNS on multi-dimensional data, and the SIFT1M dataset for ANNS on high-dimensional data.

As shown in Figure \ref{online_com}, since the SOTA method OREO is available for supporting PBF, we compare the query performance of UREM, OREO, and the original indexes over 15 sliding windows, each containing 2,000 queries~\cite{OREO}. 
The results show that both UREM and OREO achieve faster data queries compared to the original indexes, while UREM further enhances the performance of OREO by using $D_\text{opt}$ during cold start and switching. The results show that, compared with the original indexes, UREM achieves an average speedup
of 4.39$\times$.
Specifically, UREM achieves an average acceleration of 5.72$\times$ on multi-dimensional ANNS and 9.47$\times$ on high-dimensional ANNS, respectively. 
In addition, we observe that although UREM significantly outperforms the original indexes, its performance fluctuates across different sliding windows because its partial layouts are tailored to the current window. 
Nevertheless, UREM can detect such fluctuations and quickly switch to optimal partial layouts to stabilize query performance.

In addition, Figure~\ref{online_layouts}(a)-(c) presents the total time cost (including query and reorganization time) of multi- and high-dimensional indexes before and after UREM enhancement. For PBF on multi-dimensional data (Dominick dataset), results show that UREM performs fewer layout reorganizations and switches than OREO, achieves faster query times, and incurs a lower overall time cost.
For ANNS on multi-dimensional data (Forest dataset), results show that even including the time for partial layout reorganization, the total time cost of UREM remains significantly lower than the original indexes. Similarly, for ANNS on high-dimensional data (SIFT1M dataset), UREM’s total time cost remains substantially lower.

\subsubsection{Evaluation of Partial Layouts Number}\label{layoutnumber}
In Section~\ref{ExpandingPool}, we demonstrate that as the number of queries increases, the number of partial layouts in the candidate layout pool does not grow indefinitely.
Figure~\ref{online_layouts} (d) shows that as the queries expand, the growth rate of the number of layouts in the candidate layout pool gradually slows down. 
This result confirms that UREM does not lead to an uncontrolled accumulation of partial layouts under dynamic workloads.

\section{Discussion}\label{discussion}
\subsection{Adapt to Data Changes}
In many application scenarios, data may dynamically change. 
Since UREM is decoupled from the upper-level index, it preserves the original functionality of the index. 
Consequently, an index that initially supports data appending will retain this capability after UREM enhancement. 
Moreover, during layout optimization, UREM can reserve some available space in each partition to accommodate future data appending, and the newly added data can be placed into this available space.

For indexes that do not inherently support data appending, they still lack this capability after UREM enhancement under static workloads.
However, under dynamic workloads, UREM leverages partial layouts for redundant data storage, which is independent of the original index. 
As a result, UREM enables a small amount of data appending for indexes that originally do not support it, thereby extending and enriching their functionality.

\subsection{Theoretical Analysis}
\subsubsection{Complexity Analysis}
Under static workloads, the overall time complexity of UREM is $O({N_d}^2)$, and the space complexity is $O(N_d)$. 
The time cost comes from: (1) representing queries $O(N_q)$; (2) feature-based transformation $O((N_d+N_q)d)$; (3) gravity-based movement $O((N_d+N_q)^2)$; and (4) clustering $O(N_d+N_q)$. Since $N_d \gg N_q, d$, the time cost approximates as $O({N_d}^2)$. 
For space complexity, UREM requires storing (1) the data $O((N_d+N_q)d)$, and (2) clustering metadata $O(m)$. 
As a result, the total space cost is $O(N_d)$. 
These results indicate that UREM trades time cost for space cost. Since we do not focus on the overall index construction time under static workloads, the additional cost of UREM is justifiable.

Under dynamic workloads, the time cost of constructing a partial layout involves: (1) selecting a representative query point $q_{r_i}$, $O(|Q_{W_i}|)$; (2) computing distances to all data points, $O(N_d)$; (3) sorting the data, $O(N_dlogN_d)$; and (4) forming the partial layout, $O(N_l)$. Since $N_d \gg |Q_{W_i}|, O(N_l)$, the overall time cost is $O(N_dlogN_d)$. For the space cost, each partial layout space complexity is $O(N_l)$.

\subsubsection{Query Performance Analysis}
Under static workloads, UREM optimizes data layout to serve as a foundation for various indexes, thereby enabling higher query performance. As shown in Figure~\ref{offlinepartition}, UREM clusters similar data together, reducing the intra-cluster distances while increasing inter-cluster distances, resulting in lower cross-partition access rates during queries.
Experiments in Section~\ref{experimentsStatic} also demonstrate it.

Under dynamic workloads, suppose true $k$ neighbors of query $q$ lie in the partial layout $l_i$ with probability $p=N_l/N_d$. Therefore, the probability that the query accesses the entire dataset is $1 - p$.
If the query result is successfully found within $l$ by the B-tree, the cost is $\log N_l$. 
Otherwise, UREM falls back to the full dataset $D_\text{opt}$, incurring the query cost $C$, where $C$ is determined by the upper-level index. 
Therefore, the expected query cost is $
\mathbb{E}(c(q,l)) = (1-(1 - p)^k) \log N_l + (1 - p)^k C$. Since most indexes have an indexing efficiency of $\log N_d$ over the entire dataset, we have $C \approx \log N_d > \log N_l$. Therefore, $(1-(1 - p)^k)C > (1-(1 - p)^k)\log N_l$, which implies $
\mathbb{E}(c(q,l)) = (1-(1 - p)^k) \log N_l + (1 - p)^k C < C$.
Therefore, $l_i$ can enhance the query performance.

\section{Related Work}\label{relatedwork}
To improve query performance for multi- and high-dimensional data, various enhancement methods have been proposed, which can be categorized into structure-oriented enhancement methods and layout-oriented enhancement methods.

Most existing enhancement methods are structure-oriented enhancement methods. As shown in Table~\ref{Trelatedwork}, these methods construct reliable index structures by leveraging data characteristics and static workload patterns, which can be categorized into data- and query-driven methods. 
For data-driven methods, some multi-dimensional indexes~\cite{ZM, Ml-index, LISA} utilize learning models to capture data distributions. 
For high-dimensional data, some indexes~\cite{E2LSH, DB_LSH} leverage dimensionality reduction, while other methods~\cite{IVFADC} use clustering to coarsely partition the data. 
For query-driven methods, multi-dimensional indexes~\cite{Flood, Tsunami} rely on learning models to achieve query awareness, while a few high-dimensional indexes~\cite{QALSH} construct index structures regarding queries by dimensionality reduction.
Although these methods are effective, they optimistically assume that workloads are static and invariant over time, which rarely holds in practice. Subsequent studies proposed layout-oriented enhancement methods~\cite{fine-grained, pando, OREO, SAT, MTO, cracking, depa}, which stabilize query performance by dynamically optimizing data layouts. However, these studies still lack exploration of integration with higher-level indexes and remain limited to PBF queries on multi-dimensional data.

\section{Conclusion}\label{conclusion}
In this paper, we propose a unified method, called UREM, for enhancing the query performance of multi- and high-dimensional indexes in different scenarios. 
As a layout-oriented enhancement method decoupled from specific indexes, UREM offline deeply analyzes the data characteristics and query patterns under static workloads to perform layout optimization, thereby providing the upper-level index with a reliable data layout. 
Under dynamic workloads, UREM uses online information to maintain a candidate layout pool. 
It adaptively selects or generates the optimal partial layout based on continuously shifting sliding windows, ensuring stable query performance. 
Extensive experiments demonstrate that UREM can effectively improve query performance under both static and dynamic workloads.

\bibliographystyle{IEEEtran}
\bibliography{sample}

\vspace{12pt}
\color{red}

\end{document}